\documentclass[twocolumn,prl,showpacs,amsmath,amssymb]{revtex4}

\usepackage{graphicx}
\usepackage{color}
\usepackage{dcolumn}% Align table columns on decimal point
\usepackage{bm}% bold math
\include{epsf}

\begin{document}

\title{Electrical excitation of surface plasmons}

\author{Palash Bharadwaj$^1$, Alexandre Bouhelier$^2$ and Lukas Novotny$^1$}
\homepage{www.nano-optics.org}
\affiliation{$^{1}$Institute of Optics and Department of Physics and Astronomy, University of Rochester, Rochester, NY 14627, USA}
\affiliation{$^{2}$Laboratoire Interdisciplinaire Carnot de Bourgogne, CNRS UMR-5209, Universit\'e de Bourgogne, 21000 Dijon, France}

\date{\today}

\begin{abstract}
%Non-relativistic free electrons do not couple readily to light because of a large momentum mismatch. This mismatch can however be overcome by intermediate plasmon polaritons, hybrid modes of electrons and photons supported by metal surfaces. Here 
We exploit a plasmon mediated two-step momentum downconversion scheme to convert low-energy tunneling electrons into propagating photons. 
Surface plasmon polaritons (SPPs) propagating along an extended gold nanowire are excited on one end by low-energy electron tunneling and are then 
%scattered and 
converted to free-propagating photons at the other end. The separation of excitation and outcoupling proofs that tunneling electrons excite gap plasmons that subsequently couple to propagating plasmons. 
Our work shows that electron tunneling provides a non-optical, voltage-controlled and low-energy pathway for launching SPPs in nanostructures, such as plasmonic waveguides. 
%It demands metal electrodes that can be easily integrated in a planar geometry and the low-voltages required make it compatible with current complementary metal oxide semiconductor (CMOS) technology,  thus enabling electron-photon signal transduction in nanophotonic devices.  
%
%For the gold nanowire we determine an electron-to-photon conversion efficiency of $10^{-4}$. This number can be enhanced by suitably engineered plasmonic structures.  \\[-2ex]
%
%Our results hold promise for controlled electrical plasmon generation in nanophotonic devices.
\end{abstract}

\maketitle

The momentum of a non-relativistic free electron of mass $m_e$ is a factor $\sqrt{2m_ec^2/\hbar \omega}\sim500$ greater than that of a visible photon of the same energy.
% $\hbar\omega$ as the electron.  
This implies that the interaction between electrons and photons is weak 
for $\hbar \omega\ll m_ec^2$, since energy and momentum cannot be simultaneously conserved. One way to bridge this mismatch is to employ polariton modes (plasmons). They have the same energy as free space photons but arbitrarily high spatial localization (hence momentum),  and can therefore couple efficiently to electrons~\cite{garcia10}.  \\[-2ex]

The idea to use electrons to excite surface plasmon polaritons (SPPs) in metals dates back to Ritchie's landmark paper of 1957~\cite{ritchie57}. High-energy ($\sim\!30\,$keV) free electrons have since been used to excite both propagating~\cite{bashevoy06,vesseur07,cai09} and localized plasmons~\cite{nelayah07,kuttge09}.  A novel low energy route based on inelastic electron tunneling was discovered by Lambe and McCarthy in experiments with roughened metal-insulator-metal (MIM) junctions~\cite{lambe76}, ultimately leading to the observation of light emission from the scanning tunneling microscope (STM)~\cite{gimzewski88,coombs88}.  The origin of STM  photoemission  lies in the highly localized gap plasmons excited in the tip-sample cavity by inelastically tunneling electrons~\cite{johansson90,berndt91}.
%,berndt98}. 
Early experiments suggested that gap plasmons could in turn excite extended surface plasmons (SPPs), which can decay radiatively in a glass substrate~\cite{uehara92,ushioda92,takeuchi91}.  Although advances in the field of STM have enabled the control of photoemission on the atomic scale~\cite{aizpurua02,dong04,schull09,chen09b}, 
%nazin03 
the  excitation of extended plasmonic structures by tunneling electrons has not yet been addressed. So far, electrical SPP excitation schemes employed either high energy electrons~\cite{bashevoy06,vesseur07,cai09} or near-field coupling of electroluminescent excitons~\cite{walters09,koller08}.  \\[-2ex]
%However,  in our opinion a truly nanoscale, integrable, electrical yet simple method for SPP generation is still missing.

\begin{figure}[!b]
%\begin{center}
%\includegraphics[width=6cm]{figure1c.eps}
\includegraphics[width=6cm]{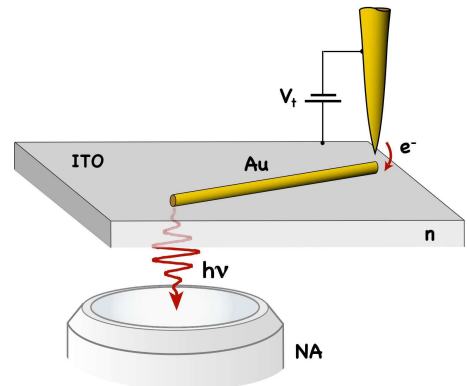}
\vspace{-3em}
\caption{Illustration of the experiment. Inelastic electron tunneling between a gold tip and a single-crystal gold nanorod gives rise to surface plasmon excitation. Locally excited surface plasmons propagate along the nanorod and scatter at the other end. Emitted photons are collected by an index-matched  NA=1.4 objective and then analyzed. 
\label{schematic}}
%\end{center}
\end{figure}
In this Letter we demonstrate a direct energy conversion route from electrons to photons, accomplished via intermediate plasmonic excitations. Specifically, we employ inelastic electron tunneling to excite localized gap plasmons in the junction of a gold tip and a monocrystalline gold nanorod. The gap plasmons then couple to propagating surface plasmon polaritons (SPPs), which in turn are converted to propagating photons by edge scattering. 
The experimental configuration, shown in Fig.~\ref{schematic}, is a combination of an inverted optical microscope and an STM. 

%~\cite{sivel92,dawson06} and leakage radiation microscopy (LRM)~\cite{hecht96c}.
%,novotny97c,bouhelier01a,drezet08}. 
A bias voltage is applied between the sample and a gold tip, whose position is held under feedback at a constant tunneling current.  
Light is collected on the substrate-side with a high numerical aperture (NA=1.4) objective, and the full angular distribution of emitted photons is recorded by imaging the back focal plane (Fourier plane) of the objective on an electron-multiplying charge-coupled device (EMCCD) camera.
% in an inverse Kretschmann configuration. 
Alternatively, the distribution of photons in real-space can be mapped by including another focusing lens in the imaging path. \\[-2ex]
%Finally, photon maps can be generated by detecting photons  point-by-point on a single-photon counting module while raster scanning the sample under the tip.\\[-2ex]

To characterize the electron-photon coupling mechanism we first investigated the electrical excitation of extended propagating SPPs on a  thermally evaporated Au film deposited on a glass  substrate (n=1.52). A film thickness of 20nm provides a compromise between SPP propagation length and film transparency for detection of leakage radiation. A chemically etched $\sim\!50$nm sharp Au tip was brought in tunneling contact to the surface (typical substrate bias V$_{\rm t}$ = 2V; tunneling current I$_{\rm t}$= 1nA) giving rise to a photon count rate of 5--10\,kcps, which corresponds to an electron-photon conversion efficiency of $\sim\!10^{-5}$. For biases of 1.9V and higher, the spectrum of the emitted light spans from 650nm to 800nm, peaking around 700nm (data not shown).  The exact spectral profile depends on the tip shape~\cite{aizpurua00}.  For a 20nm gold film and a free space wavelength of $\lambda=700$nm,  the SPP has a wavevector of $k_{spp}=1.06\,k_o$, where $k_o=2\pi/\lambda$ is the free space wavevector. Because of the finite film thickness and the supporting glass substrate, the excited SPPs decay via leakage radiation that is emitted into the Kretschmann angle  at $\theta_{{\rm K}}=43.8^{\circ}$. \\[-2ex]

\begin{figure}[!t]
%\begin{center}
%\includegraphics[width=8.5cm]{figure2b.eps}
\includegraphics[width=8.5cm]{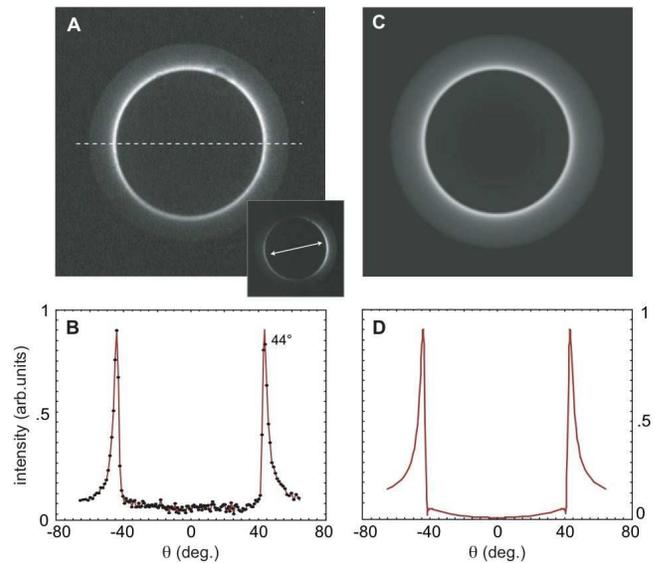}
\caption{Emission pattern (photon momentum distribution) measured in the back focal plane of the objective.
Inelastically tunneling electrons couple to propagating SPPs which decay by leakage radiation. (A) Experimental back focal plane image for a 20nm Au film showing a plasmon ring centered at $\theta\approx44^{\circ}$. V$_{\rm t}$ = 2.5V, I$_{\rm t}$= 1.5nA. Inset: Back focal plane image after passing through a polarizer oriented along the arrow in the figure.  The extinction ratio is $>\!80\%$. (B) Cross-section evaluated along the dashed line in (A).  (C) Theoretical back focal plane image for $\lambda=700\,$nm according to Eq.~(\ref{eq1}). 
%An optical apodization function of $\cos\theta$ has been used. 
(D) Theoretical cross-section through the center of (C).  
\label{fourier}}
%\end{center}
\end{figure}

To experimentally verify the electrical excitation of SPPs and their radiative decay we imaged the intensity distribution of the emitted photons in the back focal plane of the collection objective. The recorded intensity map can be directly translated into an angular radiation pattern or a momentum distribution of the emitted photons. As shown in Figs.~\ref{fourier} (a,b), the intensity distribution is rotationally symmetric and features a ring whose radius corresponds to an emission angle of $\theta\approx44^{\circ}$. This angle is larger than the critical angle of total internal reflection $\theta_{{\rm c}}=arcsin[1/n]\approx 41.1^{\circ}$ and is readily identified as the Kretschmann angle. We find that 
more than $\sim\!90\%$ of the total radiation is emitted at angles larger than $\theta_{{\rm c}}$. To analyze the spatial polarization distribution of the emitted radiation we place a polarizer behind the back focal plane of the objective. The resulting intensity distribution  is shown in the inset of Fig.~\ref{fourier} (a) for one particular polarizer orientation. The photons in the ring are almost {\em p}-polarized, as expected for leakage radiation from propagating surface plasmons.  The incomplete extinction is likely due to scattering of localized gap modes from surface roughness. \\[-2ex]
%The discrepancy between the expected and observed angular positions of the plasmon ring most likely arises from calibration uncertainties due to pixelation ($\approx\pm1^{\circ}$), as well as systematic errors caused by imperfect focussing of the Fourier plane on the CCD. 
%Finally, the ring is inhomogenously broadened due to chromatic dispersion of the SPP, which the single-wavelength estimate does not account for.

In a next step we replaced the 20nm gold film by a $\sim\!5$nm granular gold film, which does {\em not} support propagating SPPs. As expected, gap plasmons excited by inelastic electron tunneling are not able to couple to SPPs and hence no plasmon ring could be observed. Instead, the back focal plane image features a uniform intensity pattern with a discontinuity at $\theta_{{\rm c}}$ (data not shown), suggesting that the light originates from a localized but randomly oriented source distribution~\cite{lieb04}.  \\[-2ex]

The physical mechanism behind electron-excited SPP generation can be described by a two-step process. Inelastically tunneling electrons first couple to localized modes of the tip-sample junction (gap plasmons). The excited gap plasmons then couple to SPPs propagating along the gold-air interface. The excited gap plasmons can be represented by an oscillating vertical dipole ${\bf p}_{\rm z}$. In terms of the transmission coefficient $t^{\rm (p)}$ of the gold film, the in-plane momentum ($k_{\parallel}$) distribution of transmitted photons can be calculated as~\cite{novotny06b}
\begin{equation}
\hat{\bf E}(k_{\parallel}) = \frac{i \,p_z}{8 \pi^2\varepsilon_o}\;\frac{k_{\parallel}\,t^{(p)}(k_{\parallel})}{\sqrt{1- k_{\parallel}^2/k_o^2}}\, \exp\!\left[i k_o z_o \sqrt{1\!-\! k_{\parallel}^2/k_o^2}\right] {\bf n}_{\rho}\, ,
\label{eq1}
\end{equation}
where $z_o$ is the height of the dipole above the gold surface and ${\bf n}_{\rho}$ is the unit vector in radial direction parallel to the gold layer. The intensity in the back focal plane corresponds to the modulus square of $\hat{\bf E}(k_{\parallel})$. The numerical aperture of the collection objective restricts the range of spatial frequencies to $k_{\parallel} = [0\, .. \, k_o \, {\rm NA}]$. The angle $\theta$ used for the cross-sections in Fig.~\ref{fourier} is related to $k_{\parallel}$ by
$\sin[k_{\parallel}] = {\rm NA}/n$. The theoretical intensity distribution in the back focal plane is shown in Fig.~\ref{fourier} (c) for a source dipole located at $z_o=0$. For comparison with experimental data the calculated intensity distribution has been multiplied with $\cos\theta$ in order to account for the apodization of the collection objective. A cross-section through the center of the field distribution is given in Fig.~\ref{fourier} (d). The theoretical distributions are in good agreement with the experimental data shown in Figs.~\ref{fourier} (a,b), which provides strong support for the `localized gap plasmon' hypothesis. \\[-2ex]
%While our experiments demonstrate that a thin metal film is able to convert gap plasmons into propagating photons, the planar geometry is by far not the most efficient coupling structure. Optimized plasmonic structures, such as optical antennas, will be able to drastically improve the electron-photon conversion efficiency.  \\[-2ex]

\begin{figure}[!t]
%\begin{center}
%\includegraphics[width=8.2cm]{figure3b.eps}
\includegraphics[width=8.2cm]{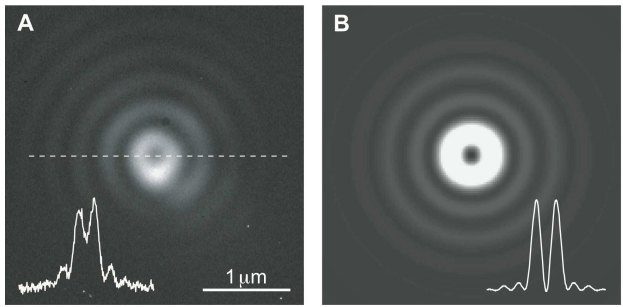}
\caption{Real space images of the tunnel junction.  (A) Experimental image and cross section recorded by refocusing the photons collected by the objective on a CCD. (B) Theoretical image corresponding to the square modulus of the Fourier transform of Eq.~(\ref{eq1}). The minimum at the center originates from the vertical dipole associated with the gap plasmon. The concentric interference rings indicate that the interfering fields have well-defined energy and momentum (leakage radiation from SPPs). 
\label{imagep}}
%\end{center}
\end{figure}

To provide further evidence for the dipole-like character of the gap plasmon we record an optical image of the tunnel junction. This is accomplished by refocusing the fields in the back focal plane of the objective on a CCD. The resulting image corresponds to the Fourier transform of Eq.~(\ref{eq1}).
%Part of the emission from the {\em z}-dipole remains uncoupled (about 30\%), and is detected as radiation beyond $\theta_{{\rm c}}$ in the Fourier plane image (c. f. Fig. 2c). 
If the dominant contribution to light emission does indeed originate from a vertical dipole,  we expect an intensity null along the optical axis, which follows from symmetry considerations.  
In Fig.~\ref{imagep} we show both the experimental image (a) and the theoretical image (b) along with cross sections through the center of the patterns. In both images we indeed find an intensity minimum at the center enclosed by concentric interference rings. These rings originate from the interference of counter propagating waves with well-defined k-vectors (leakage radiation from SPPs) on the image screen. Because of background noise, surface roughness of the film, finite detector pixel size, and  tip asymmetry, the minimum in the experimental image is not a perfect null. Nevertheless, the data agrees well with theory and provides further support for the dipole nature of the gap plasmon.\\[-2ex]

The gap plasmon can be quenched by replacing the gold tip with a material that provides sufficiently high dissipation. To demonstrate this effect we repeated our experiments with a tungsten tip. As expected, the overall intensity of the emitted radiation dropped by more than one order of magnitude (data not shown). The need for a Au-Au junction for enhanced light emission is further supported by the practice of `indenting' a tungsten tip in gold, as reported in previous studies~\cite{schull08,schull09}.\\[-2ex]

%Having confirmed the electronic excitation of SPPs on an extended gold film, we now discuss the  excitation of plasmons in confined geometries. 
For applications in nanophotonics it is desirable to have an electrically excited nanoscale photon source. For such a source to be practical, the electrical excitation region and the photon emission region need to be spatially separated, which is not the case for the extended gold film discussed so far. However, the spatial separation of excitation and emission can be accomplished by exploiting plasmons in confined geometries, such as a gold nanowire. As illustrated in Fig.~\ref{schematic}, we excite SPPs on one end of a nanowire by electron tunneling. Excited SPPs then propagate towards the other end where they are converted to propagating photons by scattering. Thus, the nanowire functions as a transmission line mediating between the electrical feed point and the optical outcoupling region.\\[-2ex]

\begin{figure}[!b]
%\begin{center}
%\includegraphics[width=8.5cm]{figure4d.eps}
\includegraphics[width=8.5cm]{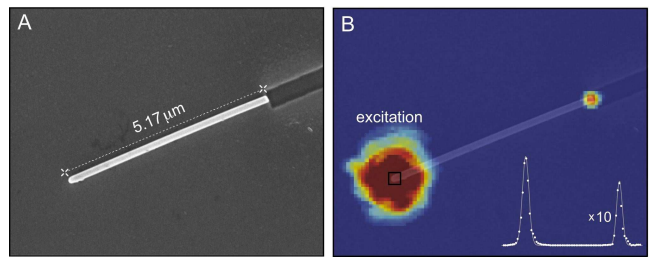}
\caption{Electrical excitation of SPPs propagating along a monocrystalline gold nanowire. (A) SEM image of a gold nanowire of radius 87 nm and length $\sim$5$\mu$m. (B) Photon emission map superimposed to the SEM image. SPPs excited at one end of the wire propagate to the other end where they are scattered and converted to propagating photons. Inset: Intensity map along the nanowire showing diffraction-limited emission peaks from both ends.
\label{wire}}
%\end{center}
\end{figure}
The monocrystalline gold nanowires used in our experiments were synthesized by wet chemistry and then dispersed on an indium tin oxide (ITO) coated glass substrate. The nanowire sample was then characterized by SEM and by optical light scattering.  To minimize SPP propagation losses and to improve the outcoupling efficiency we selected individual wires with radii in the range of 50-100 nm~\cite{shegai11}. Fig.~\ref{wire} (a) shows an SEM image of a selected nanowire of radius 87 nm that has been cut to a length of $\sim 5\,\mu$m by focused ion beam (FIB) milling. Using a translation stage we positioned the nanowire into the field of view of the objective lens and placed the STM tip over one of the wire's ends. Similar to the case of the planar gold film, electron tunneling gives rise to a gap plasmon, which can be described by an oscillating vertical dipole. The gap plasmon then couples to SPPs propagating along the gold nanowire and once the SPPs reach the other end of the nanowire they are scattered and  partially converted to free-propagating photons. Emitted photons are collected with the objective lens and then refocused on an image plane. In order to prevent propagating SPPs from decaying via leakage radiation~\cite{shegai11} the nanowire sample has been covered with index matching oil~\cite{pechou98}.  \\[-2ex]

Fig.~\ref{wire} (b) shows the direct space photon map for  the nanowire  in Fig.~\ref{wire} (a). SPPs are excited by electron tunneling at the left end of the wire. The bias voltage is V$_{\rm t}$ = 2V and the tunnel current I$_{\rm t}$= 1nA.  The map indicates that no photons are emitted along the nanowire and that a remarkable fraction of photons emerges from the wire ends. Similar results have been recorded for other wires as well. Because of the strong localization at the nanowire end, the field distribution in the backfocal plane (Fourier plane) is very broad (data not shown). For the nanowire shown in Fig.~\ref{wire} we 
%determine an electron-photon conversion efficiency in the order of $10^{-5}$. The integrated photon counts from the tunnel junction are $67,000\,s^{-1}$ whereas the total counts from the opposite end are  $5,000\,s^{-1}$.
measure $67,000$ photons $s^{-1}$ emitted from the tunnel junction whereas the total counts from the opposite end are  $5,000$ photons $s^{-1}$. Thus, the overall coupling efficiency along the nanowire is $\sim 7$\%. The efficiency improves for shorter nanowires because of lower propagation losses. Further improvement requires optimization of the coupling between gap-plasmons and propagating SPPs and optimization of the SPP outcoupling on the opposite end of the nanowire. \\[-2ex]

The nanowire serves as a proof-of-principle geometry and is by no means an optimized configuration. Engineered plasmonic structures, such as optical antennas~\cite{novotny11a}, will be able to improve the electron-photon conversion efficiency. Because of the spatial separation of SPP excitation and SPP outcoupling, the optimization can proceed in two separate steps: the optimization of the tunneling junction and  the optimization of the photon emission region. The former is a function of both the electronic and electromagnetic density of states whereas the latter constitutes a canonical antenna problem. \\[-2ex]

Our work demonstrates that  low-energy electrons can couple to photons through a cascade of events. Tunneling electrons first excite gap plasmons, which in turn couple to propagating SPPs. The latter are then converted to photons via scattering.  Thus,
%As a consequence propagating surface plasmons can be electrically excited locally by tunneling electrons.
%In conclusion, we have demonstrated direct electrical excitation of surface plasmons by employing tunneling electrons as a local source. 
electron tunneling provides a non-optical, voltage-controlled, low-energy and nanoscale pathway for launching SPPs in nanostructures. Our experiments were carried out under ambient conditions and do not require any vacuum conditions. It is important to point out the advantages of our technique over related schemes. While high energy electrons provide a highly localized source for exciting plasmons, they are impractical for on-chip implementation. Also, exciton-mediated approaches using inorganic or organic semiconductors require multi-step fabrication which increases device complexity. The method introduced here, on the other hand,  only demands metal electrodes that can be easily integrated in a planar geometry. Moreover, the low-voltages required make it completely compatible with current complementary metal oxide semiconductor (CMOS) technology,  thus enabling electron-photon signal transduction in integrated optoelectronic nanodevices.  \\

\begin{acknowledgments}
This research was funded by the U.S. Department of Energy (grant DE-FG02-01ER15204) and by the NSF (grant ECS-0651079). A. B. thanks the financial support of the NanoSci E+ program E$^2$-PLAS (ANR-08-NSCI-007). We thank Gustavo Can\c{c}ado and Hayk Harutyunyan for valuable input and fruitful discussions. 
\end{acknowledgments}

%\bibliographystyle{./apsrev}
%\bibliography{nanolib_v44}

\begin{thebibliography}{28}
\expandafter\ifx\csname natexlab\endcsname\relax\def\natexlab#1{#1}\fi
\expandafter\ifx\csname bibnamefont\endcsname\relax
  \def\bibnamefont#1{#1}\fi
\expandafter\ifx\csname bibfnamefont\endcsname\relax
  \def\bibfnamefont#1{#1}\fi
\expandafter\ifx\csname citenamefont\endcsname\relax
  \def\citenamefont#1{#1}\fi
\expandafter\ifx\csname url\endcsname\relax
  \def\url#1{\texttt{#1}}\fi
\expandafter\ifx\csname urlprefix\endcsname\relax\def\urlprefix{URL }\fi
\providecommand{\bibinfo}[2]{#2}
\providecommand{\eprint}[2][]{\url{#2}}

\bibitem[{\citenamefont{de~Abaj\'{o}}(2010)}]{garcia10}
\bibinfo{author}{\bibfnamefont{F.~J.~G.} \bibnamefont{de~Abaj\'{o}}},
  \bibinfo{journal}{Rev. of Mod. Phys.} \textbf{\bibinfo{volume}{82}},
  \bibinfo{pages}{209} (\bibinfo{year}{2010}).

\bibitem[{\citenamefont{Ritchie}(1957)}]{ritchie57}
\bibinfo{author}{\bibfnamefont{R.~H.} \bibnamefont{Ritchie}},
  \bibinfo{journal}{Phys. Rev.} \textbf{\bibinfo{volume}{106}},
  \bibinfo{pages}{874} (\bibinfo{year}{1957}).

\bibitem[{\citenamefont{Bashevoy et~al.}(2006)\citenamefont{Bashevoy, Jonsson,
  Krasavin, Zheludev, Chen, and Stockman}}]{bashevoy06}
\bibinfo{author}{\bibfnamefont{M.~V.} \bibnamefont{Bashevoy}} {\em et al.},
%  \bibinfo{author}{\bibfnamefont{F.}~\bibnamefont{Jonsson}},
%  \bibinfo{author}{\bibfnamefont{A.~V.} \bibnamefont{Krasavin}},
%  \bibinfo{author}{\bibfnamefont{N.~I.} \bibnamefont{Zheludev}},
%  \bibinfo{author}{\bibfnamefont{Y.}~\bibnamefont{Chen}}, \bibnamefont{and}
%  \bibinfo{author}{\bibfnamefont{M.~I.} \bibnamefont{Stockman}},
  \bibinfo{journal}{Nano Lett.} \textbf{\bibinfo{volume}{6}},
  \bibinfo{pages}{1113} (\bibinfo{year}{2006}).

\bibitem[{\citenamefont{Vesseur et~al.}(2007)\citenamefont{Vesseur, de~Waele,
  Kuttge, and Polman}}]{vesseur07}
\bibinfo{author}{\bibfnamefont{E.~J.~R.} \bibnamefont{Vesseur}}  {\em et al.},
%  \bibinfo{author}{\bibfnamefont{R.}~\bibnamefont{de~Waele}},
%  \bibinfo{author}{\bibfnamefont{M.}~\bibnamefont{Kuttge}}, \bibnamefont{and}
%  \bibinfo{author}{\bibfnamefont{A.}~\bibnamefont{Polman}},
  \bibinfo{journal}{Nano Lett.} \textbf{\bibinfo{volume}{7}},
  \bibinfo{pages}{2843} (\bibinfo{year}{2007}).

\bibitem[{\citenamefont{Cai et~al.}(2009)\citenamefont{Cai, Sainidou, Xu,
  Polman, and de~Abajo}}]{cai09}
\bibinfo{author}{\bibfnamefont{W.}~\bibnamefont{Cai}} {\em et al.},
%  \bibinfo{author}{\bibfnamefont{R.}~\bibnamefont{Sainidou}},
%  \bibinfo{author}{\bibfnamefont{J.}~\bibnamefont{Xu}},
%  \bibinfo{author}{\bibfnamefont{A.}~\bibnamefont{Polman}}, \bibnamefont{and}
%  \bibinfo{author}{\bibfnamefont{F.~J.~G.} \bibnamefont{de~Abajo}},
  \bibinfo{journal}{Nano Lett.} \textbf{\bibinfo{volume}{9}},
  \bibinfo{pages}{1176} (\bibinfo{year}{2009}).

\bibitem[{\citenamefont{Nelayah et~al.}(2007)\citenamefont{Nelayah, Kociak,
  St{\'e}phan, de~Abajo, Tenc{\'e}, Henrard, Taverna, Pastoriza-Santos,
  Liz-Marz{\'a}n, and Colliex}}]{nelayah07}
\bibinfo{author}{\bibfnamefont{J.}~\bibnamefont{Nelayah}}  {\em et al.},
%  \bibinfo{author}{\bibfnamefont{M.}~\bibnamefont{Kociak}},
 % \bibinfo{author}{\bibfnamefont{O.}~\bibnamefont{St{\'e}phan}},
%  \bibinfo{author}{\bibfnamefont{F.~J.~G.} \bibnamefont{de~Abajo}},
%  \bibinfo{author}{\bibfnamefont{M.}~\bibnamefont{Tenc{\'e}}},
%  \bibinfo{author}{\bibfnamefont{L.}~\bibnamefont{Henrard}},
%  \bibinfo{author}{\bibfnamefont{D.}~\bibnamefont{Taverna}},
%  \bibinfo{author}{\bibfnamefont{I.}~\bibnamefont{Pastoriza-Santos}},
%  \bibinfo{author}{\bibfnamefont{L.~M.} \bibnamefont{Liz-Marz{\'a}n}},
%  \bibnamefont{and} \bibinfo{author}{\bibfnamefont{C.}~\bibnamefont{Colliex}},
  \bibinfo{journal}{Nature Phys.} \textbf{\bibinfo{volume}{3}},
  \bibinfo{pages}{348} (\bibinfo{year}{2007}).

\bibitem[{\citenamefont{Kuttge et~al.}(2009)\citenamefont{Kuttge, Vesseur, and
  Polman}}]{kuttge09}
\bibinfo{author}{\bibfnamefont{M.}~\bibnamefont{Kuttge}},
  \bibinfo{author}{\bibfnamefont{E.~J.~R.} \bibnamefont{Vesseur}},
  \bibnamefont{and} \bibinfo{author}{\bibfnamefont{A.}~\bibnamefont{Polman}},
  \bibinfo{journal}{Appl. Phys. Lett.} \textbf{\bibinfo{volume}{94}},
  \bibinfo{pages}{183104} (\bibinfo{year}{2009}).

\bibitem[{\citenamefont{Lambe and McCarthy}(1976)}]{lambe76}
\bibinfo{author}{\bibfnamefont{J.}~\bibnamefont{Lambe}} \bibnamefont{and}
  \bibinfo{author}{\bibfnamefont{S.~L.} \bibnamefont{McCarthy}},
  \bibinfo{journal}{Phys. Rev. Lett.} \textbf{\bibinfo{volume}{37}},
  \bibinfo{pages}{923} (\bibinfo{year}{1976}).

\bibitem[{\citenamefont{Gimzewski et~al.}(1988)\citenamefont{Gimzewski, Reihl,
  Coombs, and Schlittler}}]{gimzewski88}
\bibinfo{author}{\bibfnamefont{J.~K.} \bibnamefont{Gimzewski}},
  \bibinfo{author}{\bibfnamefont{B.}~\bibnamefont{Reihl}},
  \bibinfo{author}{\bibfnamefont{J.~H.} \bibnamefont{Coombs}},
  \bibnamefont{and} \bibinfo{author}{\bibfnamefont{R.~R.}
  \bibnamefont{Schlittler}}, \bibinfo{journal}{Z. Phys. B}
  \textbf{\bibinfo{volume}{72}}, \bibinfo{pages}{497} (\bibinfo{year}{1988}).

\bibitem[{\citenamefont{Coombs et~al.}(1988)\citenamefont{Coombs, Gimzewski,
  Reihl, and Sass}}]{coombs88}
\bibinfo{author}{\bibfnamefont{J.}~\bibnamefont{Coombs}},
  \bibinfo{author}{\bibfnamefont{J.}~\bibnamefont{Gimzewski}},
  \bibinfo{author}{\bibfnamefont{B.}~\bibnamefont{Reihl}}, \bibnamefont{and}
  \bibinfo{author}{\bibfnamefont{J.}~\bibnamefont{Sass}}, \bibinfo{journal}{J.
  Microsc.} \textbf{\bibinfo{volume}{152}}, \bibinfo{pages}{325}
  (\bibinfo{year}{1988}).

\bibitem[{\citenamefont{Johansson et~al.}(1990)\citenamefont{Johansson,
  Monreal, and Apell}}]{johansson90}
\bibinfo{author}{\bibfnamefont{P.}~\bibnamefont{Johansson}},
  \bibinfo{author}{\bibfnamefont{R.}~\bibnamefont{Monreal}}, \bibnamefont{and}
  \bibinfo{author}{\bibfnamefont{P.}~\bibnamefont{Apell}},
  \bibinfo{journal}{Phys. Rev. B} \textbf{\bibinfo{volume}{42}},
  \bibinfo{pages}{9210} (\bibinfo{year}{1990}).

\bibitem[{\citenamefont{Berndt et~al.}(1991)\citenamefont{Berndt, Gimzewski,
  and Johansson}}]{berndt91}
\bibinfo{author}{\bibfnamefont{R.}~\bibnamefont{Berndt}},
  \bibinfo{author}{\bibfnamefont{J.~K.} \bibnamefont{Gimzewski}},
  \bibnamefont{and}
  \bibinfo{author}{\bibfnamefont{P.}~\bibnamefont{Johansson}},
  \bibinfo{journal}{Phys. Rev. Lett.} \textbf{\bibinfo{volume}{67}},
  \bibinfo{pages}{3796} (\bibinfo{year}{1991}).

\bibitem[{\citenamefont{Uehara et~al.}(1992)\citenamefont{Uehara, Kimura,
  Ushioda, and Takeuchi}}]{uehara92}
\bibinfo{author}{\bibfnamefont{Y.}~\bibnamefont{Uehara}},
  \bibinfo{author}{\bibfnamefont{Y.}~\bibnamefont{Kimura}},
  \bibinfo{author}{\bibfnamefont{S.}~\bibnamefont{Ushioda}}, \bibnamefont{and}
  \bibinfo{author}{\bibfnamefont{K.}~\bibnamefont{Takeuchi}},
  \bibinfo{journal}{Jap. J. of Appl. Phys.} \textbf{\bibinfo{volume}{31}},
  \bibinfo{pages}{2465} (\bibinfo{year}{1992}).

\bibitem[{\citenamefont{Ushioda et~al.}(1992)\citenamefont{Ushioda, Uehara, and
  Kuwahara}}]{ushioda92}
\bibinfo{author}{\bibfnamefont{S.}~\bibnamefont{Ushioda}},
  \bibinfo{author}{\bibfnamefont{Y.}~\bibnamefont{Uehara}}, \bibnamefont{and}
  \bibinfo{author}{\bibfnamefont{M.}~\bibnamefont{Kuwahara}},
  \bibinfo{journal}{Appl. Surf. Sci.} \textbf{\bibinfo{volume}{60/61}},
  \bibinfo{pages}{448} (\bibinfo{year}{1992}).

\bibitem[{\citenamefont{Takeuchi et~al.}(1991)\citenamefont{Takeuchi, Uehara,
  Ushioda, and Morita}}]{takeuchi91}
\bibinfo{author}{\bibfnamefont{K.}~\bibnamefont{Takeuchi}},
  \bibinfo{author}{\bibfnamefont{Y.}~\bibnamefont{Uehara}},
  \bibinfo{author}{\bibfnamefont{S.}~\bibnamefont{Ushioda}}, \bibnamefont{and}
  \bibinfo{author}{\bibfnamefont{S.}~\bibnamefont{Morita}},
  \bibinfo{journal}{J. Vac. Sci. and Technol. B} \textbf{\bibinfo{volume}{9}},
  \bibinfo{pages}{557} (\bibinfo{year}{1991}).

\bibitem[{\citenamefont{Aizpurua et~al.}(2002)\citenamefont{Aizpurua, Hoffmann,
  Apell, and Berndt}}]{aizpurua02}
\bibinfo{author}{\bibfnamefont{J.}~\bibnamefont{Aizpurua}},
  \bibinfo{author}{\bibfnamefont{G.}~\bibnamefont{Hoffmann}},
  \bibinfo{author}{\bibfnamefont{S.~P.} \bibnamefont{Apell}}, \bibnamefont{and}
  \bibinfo{author}{\bibfnamefont{R.}~\bibnamefont{Berndt}},
  \bibinfo{journal}{Phys. Rev. Lett.} \textbf{\bibinfo{volume}{89}},
  \bibinfo{pages}{156803} (\bibinfo{year}{2002}).

\bibitem[{\citenamefont{Dong et~al.}(2004)\citenamefont{Dong, Guo, Trifonov,
  Dorozhkin, Miki, Kimura, Yokoyama, and Mashiko}}]{dong04}
\bibinfo{author}{\bibfnamefont{Z.~C.} \bibnamefont{Dong}}  {\em et al.},
%  \bibinfo{author}{\bibfnamefont{X.~L.} \bibnamefont{Guo}},
%  \bibinfo{author}{\bibfnamefont{A.~S.} \bibnamefont{Trifonov}},
%  \bibinfo{author}{\bibfnamefont{P.~S.} \bibnamefont{Dorozhkin}},
%  \bibinfo{author}{\bibfnamefont{K.}~\bibnamefont{Miki}},
%  \bibinfo{author}{\bibfnamefont{K.}~\bibnamefont{Kimura}},
%  \bibinfo{author}{\bibfnamefont{S.}~\bibnamefont{Yokoyama}}, \bibnamefont{and}
%  \bibinfo{author}{\bibfnamefont{S.}~\bibnamefont{Mashiko}},
  \bibinfo{journal}{Phys. Rev. Lett.} \textbf{\bibinfo{volume}{92}},
  \bibinfo{pages}{86801} (\bibinfo{year}{2004}).

\bibitem[{\citenamefont{Schull et~al.}(2009)\citenamefont{Schull, N{\'e}el,
  Johansson, and Berndt}}]{schull09}
\bibinfo{author}{\bibfnamefont{G.}~\bibnamefont{Schull}},
  \bibinfo{author}{\bibfnamefont{N.}~\bibnamefont{N{\'e}el}},
  \bibinfo{author}{\bibfnamefont{P.}~\bibnamefont{Johansson}},
  \bibnamefont{and} \bibinfo{author}{\bibfnamefont{R.}~\bibnamefont{Berndt}},
  \bibinfo{journal}{Phys. Rev. Lett.} \textbf{\bibinfo{volume}{102}},
  \bibinfo{pages}{057401} (\bibinfo{year}{2009}).

\bibitem[{\citenamefont{Chen et~al.}(2009)\citenamefont{Chen, Bobisch, and
  Ho}}]{chen09b}
\bibinfo{author}{\bibfnamefont{C.}~\bibnamefont{Chen}},
  \bibinfo{author}{\bibfnamefont{C.~A.} \bibnamefont{Bobisch}},
  \bibnamefont{and} \bibinfo{author}{\bibfnamefont{W.}~\bibnamefont{Ho}},
  \bibinfo{journal}{Science} \textbf{\bibinfo{volume}{325}},
  \bibinfo{pages}{981} (\bibinfo{year}{2009}).

\bibitem[{\citenamefont{Walters et~al.}(2009)\citenamefont{Walters, van Loon,
  Brunets, Schmitz, and Polman}}]{walters09}
\bibinfo{author}{\bibfnamefont{R.~J.} \bibnamefont{Walters}} {\em et al.},
%  \bibinfo{author}{\bibfnamefont{R.~V.~A.} \bibnamefont{van Loon}},
%  \bibinfo{author}{\bibfnamefont{I.}~\bibnamefont{Brunets}},
%  \bibinfo{author}{\bibfnamefont{J.}~\bibnamefont{Schmitz}}, \bibnamefont{and}
%  \bibinfo{author}{\bibfnamefont{A.}~\bibnamefont{Polman}},
  \bibinfo{journal}{Nature Mat.} \textbf{\bibinfo{volume}{9}},
  \bibinfo{pages}{21} (\bibinfo{year}{2009}).

\bibitem[{\citenamefont{Koller et~al.}(2008)\citenamefont{Koller, Hohenau,
  Ditlbacher, Galler, Reil, Aussenegg, Leitner, Lis, and Krenn}}]{koller08}
\bibinfo{author}{\bibfnamefont{D.~M.} \bibnamefont{Koller}} {\em et al.},
%  \bibinfo{author}{\bibfnamefont{A.}~\bibnamefont{Hohenau}},
%  \bibinfo{author}{\bibfnamefont{H.}~\bibnamefont{Ditlbacher}},
%  \bibinfo{author}{\bibfnamefont{N.}~\bibnamefont{Galler}},
%  \bibinfo{author}{\bibfnamefont{F.}~\bibnamefont{Reil}},
%  \bibinfo{author}{\bibfnamefont{F.~R.} \bibnamefont{Aussenegg}},
%  \bibinfo{author}{\bibfnamefont{A.}~\bibnamefont{Leitner}},
%  \bibinfo{author}{\bibfnamefont{E.~J.~W.} \bibnamefont{Lis}},
%  \bibnamefont{and} \bibinfo{author}{\bibfnamefont{J.~R.} \bibnamefont{Krenn}},
  \bibinfo{journal}{Nature Photon.} \textbf{\bibinfo{volume}{2}},
  \bibinfo{pages}{684} (\bibinfo{year}{2008}).

\bibitem[{\citenamefont{Aizpurua et~al.}(2000)\citenamefont{Aizpurua, Apell,
  and Berndt}}]{aizpurua00}
\bibinfo{author}{\bibfnamefont{J.}~\bibnamefont{Aizpurua}},
  \bibinfo{author}{\bibfnamefont{S.~P.} \bibnamefont{Apell}}, \bibnamefont{and}
  \bibinfo{author}{\bibfnamefont{R.}~\bibnamefont{Berndt}},
  \bibinfo{journal}{Phys. Rev. B} \textbf{\bibinfo{volume}{62}},
  \bibinfo{pages}{2065} (\bibinfo{year}{2000}).

\bibitem[{\citenamefont{Lieb et~al.}(2004)\citenamefont{Lieb, Zavislan, and
  Novotny}}]{lieb04}
\bibinfo{author}{\bibfnamefont{M.~A.} \bibnamefont{Lieb}},
  \bibinfo{author}{\bibfnamefont{J.}~\bibnamefont{Zavislan}}, \bibnamefont{and}
  \bibinfo{author}{\bibfnamefont{L.}~\bibnamefont{Novotny}},
  \bibinfo{journal}{J. Opt. Soc. Am. B} \textbf{\bibinfo{volume}{21}},
  \bibinfo{pages}{1210} (\bibinfo{year}{2004}).

\bibitem[{\citenamefont{Novotny and Hecht}(2006)}]{novotny06b}
\bibinfo{author}{\bibfnamefont{L.}~\bibnamefont{Novotny}} \bibnamefont{and}
  \bibinfo{author}{\bibfnamefont{B.}~\bibnamefont{Hecht}},
  \emph{\bibinfo{title}{Principles of Nano-Optics}}
  (\bibinfo{publisher}{Cambridge University Press},
  \bibinfo{address}{Cambridge}, \bibinfo{year}{2006}).

\bibitem[{\citenamefont{Schull et~al.}(2008)\citenamefont{Schull, Becker, and
  Berndt}}]{schull08}
\bibinfo{author}{\bibfnamefont{G.}~\bibnamefont{Schull}},
  \bibinfo{author}{\bibfnamefont{M.}~\bibnamefont{Becker}}, \bibnamefont{and}
  \bibinfo{author}{\bibfnamefont{R.}~\bibnamefont{Berndt}},
  \bibinfo{journal}{Phys. Rev. Lett.} \textbf{\bibinfo{volume}{101}},
  \bibinfo{pages}{136801} (\bibinfo{year}{2008}).

\bibitem[{\citenamefont{Shegai et~al.}(2011)\citenamefont{Shegai, Miljkovi,
  Bao, Xu, Nordlander, Johansson, and K{\"a}ll}}]{shegai11}
\bibinfo{author}{\bibfnamefont{T.}~\bibnamefont{Shegai}} {\em et al.},
%  \bibinfo{author}{\bibfnamefont{V.~D.} \bibnamefont{Miljkovi}},
%  \bibinfo{author}{\bibfnamefont{K.}~\bibnamefont{Bao}},
%  \bibinfo{author}{\bibfnamefont{H.}~\bibnamefont{Xu}},
%  \bibinfo{author}{\bibfnamefont{P.}~\bibnamefont{Nordlander}},
%  \bibinfo{author}{\bibfnamefont{P.}~\bibnamefont{Johansson}},
%  \bibnamefont{and} \bibinfo{author}{\bibfnamefont{M.}~\bibnamefont{K{\"a}ll}},
  \bibinfo{journal}{Nano Lett.} \textbf{\bibinfo{volume}{11}},
  \bibinfo{pages}{706} (\bibinfo{year}{2011}).

\bibitem[{\citenamefont{Pechou et~al.}(1998)\citenamefont{Pechou, Coratger,
  Girardin, Ajustron, and Beauvillain}}]{pechou98}
\bibinfo{author}{\bibfnamefont{R.}~\bibnamefont{Pechou}} {\em et al.},
%  \bibinfo{author}{\bibfnamefont{R.}~\bibnamefont{Coratger}},
%  \bibinfo{author}{\bibfnamefont{C.}~\bibnamefont{Girardin}},
%  \bibinfo{author}{\bibfnamefont{F.}~\bibnamefont{Ajustron}}, \bibnamefont{and}
%  \bibinfo{author}{\bibfnamefont{J.}~\bibnamefont{Beauvillain}},
  \bibinfo{journal}{Eur. Phys. J. Appl. Phys.} \textbf{\bibinfo{volume}{2}},
  \bibinfo{pages}{135} (\bibinfo{year}{1998}).

\bibitem[{\citenamefont{Novotny and van Hulst}(2011)}]{novotny11a}
\bibinfo{author}{\bibfnamefont{L.}~\bibnamefont{Novotny}} \bibnamefont{and}
  \bibinfo{author}{\bibfnamefont{N.~F.} \bibnamefont{van Hulst}},
  \bibinfo{journal}{Nature Phot.} \textbf{\bibinfo{volume}{5}},
  \bibinfo{pages}{83} (\bibinfo{year}{2011}).

\end{thebibliography}

\end{document}